\documentstyle[10pt,epsf,epsfig,dp_delphititle]{dp_delphi}
\makeindex
\def\DpPaperGroup{EP}
\def\DpPaperRef{2000-059}
\def\DpDate{3 May 2000}
\def\DpAuthors{DELPHI Collaboration}
\def\DpSubmit{(Submitted to Physics Letters B)}
\def\DpTitle{{{Search for SUSY with \boldmath$R$--parity violating
       \boldmath $\rm{LL} \bar{\rm E}$
       couplings at $\sqrt{s}$~=~189~GeV}}}
\def\DpComment{ }
\def\DpEMail{ }

\newcommand{\XOI} {$\widetilde{\chi}_1^0$}

\newcommand{\XOII} {$\widetilde{\chi}_2^0$}

\newcommand{\XOi} {$\widetilde{\chi}_i^0$}
\newcommand{\XOj} {$\widetilde{\chi}_j^0$}

\newcommand{\XPI}{$\widetilde{\chi}_1^+$}

\newcommand{\XMI}{$\widetilde{\chi}_1^-$}

\newcommand{\XPMI}{$\widetilde{\chi}_1^{\pm}$}
\newcommand{\XPk}{$\widetilde{\chi}_k^+$}
\newcommand{\XMl}{$\widetilde{\chi}_l^-$}

\newcommand{\SNU}{$\widetilde{\nu}$}

\newcommand{\slep} {\mbox{$ \tilde{\ell} $}}

\newcommand{\lum}{$\cal L$}

\newcommand{\tanb}{tan$\beta$}
\newcommand{\Labb}{$\lambda_{122}$}
\newcommand{\Lacc}{$\lambda_{133}$}
\newcommand{\Lijk}{$\lambda_{ijk}$}

\newcommand{\LiLjEk} {$ {\rm{L}}_i {\rm{L}}_j {\rm\bar{E}}_k $}

\newcommand{\LLE} {$ {\rm {LL \bar E}}$}

\newcommand{\Pt}{\mbox{p$_{\rm t}$}}
\newcommand{\Ptm}{\mbox{$\not \! {\rm p_{\rm t}}$}}
\newcommand{\Ra}{$\rightarrow$\ }

\newcommand{\WW} {\mbox{$ {\mathrm W}^+{\mathrm W}^- $}}
\newcommand{\ZZ} {\mbox{$ {\mathrm Z}{\mathrm Z} $}}

\newcommand{\Zg} {\mbox{$ {\mathrm f \bar{\mathrm f}} \gamma $}}

\newcommand{\ee} {\mbox{$ {\mathrm e}^+ {\mathrm e}^- $}}

\newcommand{\GeV} {\mbox{$ {\mathrm{GeV}} $}}
\newcommand{\GeVc} {\mbox{$ {\mathrm{GeV}}/c $}}
\newcommand{\GeVcc} {\mbox{$ {\mathrm{GeV}}/c^2 $}}

\newcommand{\Wev} {\mbox{$ {\mathrm{W e}} \nu_{\mathrm e} $}}

\newcommand{\Zee} {\mbox{$ {\mathrm Z} \ee $}}

\newcommand{\sel} {\mbox{$ \tilde{{\rm e}} $}}

\newcommand{\smu} {\mbox{$ \tilde{\mu} $}}

\newcommand{\stau} {\mbox{$ \tilde{\tau} $}}

\newcommand{\staur} {\mbox{$ \widetilde{\tau}_R $}}
\newcommand{\snu} {\mbox{$ \tilde{\nu} $}}
\newcommand{\snue} {\mbox{$ \widetilde{\nu}_{{\rm e}} $}}
\newcommand{\snum} {\mbox{$ \widetilde{\nu}_{\mu} $}}
\newcommand{\snut} {\mbox{$ \widetilde{\nu}_{\tau} $}}
\newcommand{\asnu} {\mbox{$ \tilde{\bar \nu} $}}
\newcommand{\asnue} {\mbox{$ \widetilde{\bar \nu}_{{\rm e}} $}}
\newcommand{\asnum} {\mbox{$ \widetilde{\bar \nu}_{\mu} $}}
\newcommand{\asnut} {\mbox{$ \widetilde{\bar \nu}_{\tau} $}}

\newcommand{\mydeg} {\mbox{$ ^\circ $}}

\newcommand {\rpv} {$\not \! {R}_{p}$}

\def \missEt {\ifmmode{/\mkern-11mu E_t}\else{${/\mkern-11mu E_t}$}\fi}
\def \missE {\ifmmode{/\mkern-11mu E}\else{${/\mkern-11mu E}$}\fi}

\newenvironment{malist}
{\begin{list}{\tiny{$\bullet$}}
{\setlength{\parsep}{0pt}
\addtolength{\leftmargin}{0mm}
\setlength{\topsep}{0pt}
\setlength{\itemsep}{0pt}}}{\end{list}}

\def\ap#1#2#3 {{\em Ann. Phys. (NY)} {\bf#1} (#2) #3.}
\def\apj#1#2#3 {{\em Astrophys. J.} {\bf#1} (#2) #3.}
\def\apjl#1#2#3 {{\em Astrophys. J. Lett.} {\bf#1} (#2) #3.}
\def\app#1#2#3 {{\em Acta. Phys. Pol.} {\bf#1} (#2) #3.}
\def\ar#1#2#3 {{\em Ann. Rev. Nucl. Part. Sci.} {\bf#1} (#2) #3.}
\def\cpc#1#2#3 {{\em Computer Phys. Comm.} {\bf#1} (#2) #3.}
\def\err#1#2#3 {{\it Erratum} {\bf#1} (#2) #3.}
\def\epj#1#2#3 {\mbox{{\em Eur. Phys. J.~}{\bf#1} (#2) #3}}
\def\ib#1#2#3 {{\it ibid.} {\bf#1} (#2) #3.}
\def\jmp#1#2#3 {{\em J. Math. Phys.} {\bf#1} (#2) #3.}
\def\ijmp#1#2#3 {{\em Int. J. Mod. Phys.} {\bf#1} (#2) #3.}
\def\jetp#1#2#3 {{\em JETP Lett.} {\bf#1} (#2) #3.}
\def\jpg#1#2#3 {{\em J. Phys. G.} {\bf#1} (#2) #3.}
\def\mpl#1#2#3 {{\em Mod. Phys. Lett.} {\bf#1} (#2) #3.}
\def\nat#1#2#3 {{\em Nature (London)} {\bf#1} (#2) #3.}
\def\nc#1#2#3 {{\em Nuovo Cim.} {\bf#1} (#2) #3.}
\def\nim#1#2#3 {{\em Nucl. Instr. Meth.} {\bf#1} (#2) #3.}
\def\np#1#2#3 {{\em Nucl. Phys.} {\bf#1} (#2) #3.}
\def\npsup#1#2#3#4 {{\em Nucl. Phys.} {\bf #1} {\em (Proc. Suppl.)}
{\bf#2} (#3) #4.}
\def\pcps#1#2#3 {{\em Proc. Cam. Phil. Soc.} {\bf#1} (#2) #3.}
\def\pl#1#2#3 {{\em Phys. Lett.} {\bf#1} (#2) #3.}
\def\prep#1#2#3 {{\em Phys. Rep.} {\bf#1} (#2) #3.}
\def\prev#1#2#3 {{\em Phys. Rev.} {\bf#1} (#2) #3.}
\def\prl#1#2#3 {{\em Phys. Rev. Lett.} {\bf#1} (#2) #3.}
\def\prs#1#2#3 {{\em Proc. Roy. Soc.} {\bf#1} (#2) #3.}
\def\ptp#1#2#3 {{\em Prog. Th. Phys.} {\bf#1} (#2) #3.}
\def\ps#1#2#3 {{\em Physica Scripta} {\bf#1} (#2) #3.}
\def\rmp#1#2#3 {{\em Rev. Mod. Phys.} {\bf#1} (#2) #3.}
\def\rpp#1#2#3 {{\em Rep. Prog. Phys.} {\bf#1} (#2) #3.}
\def\sjnp#1#2#3 {{\em Sov. J. Nucl. Phys.} {\bf#1} (#2) #3.}
\def\spj#1#2#3 {{\em Sov. Phys. JEPT} {\bf#1} (#2) #3.}
\def\spu#1#2#3 {{\em Sov. Phys.-Usp.} {\bf#1} (#2) #3.}
\def\zp#1#2#3 {{\em Zeit. Phys.} {\bf#1} (#2) #3.}
\def\delnote#1#2#3#4 {{DELPHI} {#1}-{#2}~{\sc #3}~#4.}
\def\opalnote#1#2 {{OPAL} {\it#1} {#2}}
\def\alnote#1#2#3#4 {{ALEPH} {#1}-{#2}~{\sc #3}~#4}
\begin{document}
\makeatletter
\newcount\@tempcntc
\def\@citex[#1]#2{\if@filesw\immediate\write\@auxout{\string\citation{#2}}\fi
\@tempcnta\z@\@tempcntb\m@ne\def\@citea{}\@cite{\@for\@citeb:=#2\do
{\@ifundefined
{b@\@citeb}{\@citeo\@tempcntb\m@ne\@citea\def\@citea{,}{\bf ?}\@warning
{Citation `\@citeb' on page \thepage \space undefined}}%
{\setbox\z@\hbox{\global\@tempcntc0\csname b@\@citeb\endcsname\relax}%
\ifnum\@tempcntc=\z@ \@citeo\@tempcntb\m@ne
\@citea\def\@citea{,}\hbox{\csname b@\@citeb\endcsname}%
\else
\advance\@tempcntb\@ne
\ifnum\@tempcntb=\@tempcntc
\else\advance\@tempcntb\m@ne\@citeo
\@tempcnta\@tempcntc\@tempcntb\@tempcntc\fi\fi}}\@citeo}{#1}}
\def\@citeo{\ifnum\@tempcnta>\@tempcntb\else\@citea\def\@citea{,}%
\ifnum\@tempcnta=\@tempcntb\the\@tempcnta\else
{\advance\@tempcnta\@ne\ifnum\@tempcnta=\@tempcntb \else \def\@citea{--}\fi
\advance\@tempcnta\m@ne\the\@tempcnta\@citea\the\@tempcntb}\fi\fi}
\makeatother
\begin{titlepage}
\pagenumbering{roman}
\CERNpreprint{\DpPaperGroup}{\DpPaperRef} 
\date{{\small\DpDate}} 
\title{\DpTitle} 
\address{\DpAuthors} 
\begin{shortabs} 
\noindent
%
\noindent
Searches for pair production of supersymmetric particles
under the assumption that $R$--parity is not conserved are presented,
based on data recorded by the DELPHI detector in 1998
from e$^{+}$e$^{-}$ collisions at a centre-of-mass energy of
189~GeV. Only one $R$--parity violating  $\rm{LL} \bar{\rm E}$
term (i.e. one $\lambda$ coupling), which couples 
\mbox{scalar} leptons to  leptons,
is considered to be dominant at a time.  
Moreover, it is assumed that the strength of the $R$--parity 
violating couplings is such that the lifetimes can be neglected. 
The search for pair production of neutralinos, charginos and sleptons
has been performed for both direct $R$--parity violating decays
and indirect cascade decays.
The results are in agreement with Standard Model expectations, and
are used to update the constraints on the MSSM parameter
values and the mass limits previously derived at \mbox{$\sqrt{s}$~=~183~GeV.}\\
The present 95\% C.L. limits on supersymmetric particle masses are:
\begin{malist}
\item $m_{\tilde{\chi}^0}>~$30~\GeVcc\ and $m_{\tilde{\chi}^\pm}>$~94~\GeVcc;
\item $m_{\tilde{\nu}}>$~76.5~\GeVcc\ (direct and indirect decays);
\item $m_{\tilde{\ell}_R}>$~83~\GeVcc\ (indirect decay only).
\end{malist}

\end{shortabs}
\vfill
\begin{center}
\DpSubmit \ \\ 
\DpComment \ \\
\DpEMail \ \\
\end{center}
\vfill
\clearpage
\headsep 10.0pt
\addtolength{\textheight}{10mm}
\addtolength{\footskip}{-5mm}
\begingroup
%
\newcommand{\DpName}[2]{\hbox{#1$^{\ref{#2}}$},\hfill}
\newcommand{\DpNameTwo}[3]{\hbox{#1$^{\ref{#2},\ref{#3}}$},\hfill}
\newcommand{\DpNameThree}[4]{\hbox{#1$^{\ref{#2},\ref{#3},\ref{#4}}$},\hfill}
\newskip\Bigfill \Bigfill = 0pt plus 1000fill
\newcommand{\DpNameLast}[2]{\hbox{#1$^{\ref{#2}}$}\hspace{\Bigfill}}
%
\footnotesize
\noindent
\DpName{P.Abreu}{LIP}
\DpName{W.Adam}{VIENNA}
\DpName{T.Adye}{RAL}
\DpName{P.Adzic}{DEMOKRITOS}
\DpName{Z.Albrecht}{KARLSRUHE}
\DpName{T.Alderweireld}{AIM}
\DpName{G.D.Alekseev}{JINR}
\DpName{R.Alemany}{VALENCIA}
\DpName{T.Allmendinger}{KARLSRUHE}
\DpName{P.P.Allport}{LIVERPOOL}
% ...
\normalsize
\endgroup
\titlefoot{Department of Physics and Astronomy, Iowa State
     University, Ames IA 50011-3160, USA
    \label{AMES}}
\titlefoot{Physics Department, Univ. Instelling Antwerpen,
     Universiteitsplein 1, B-2610 Antwerpen, Belgium \\
     \indent~~and IIHE, ULB-VUB,
     Pleinlaan 2, B-1050 Brussels, Belgium \\
     \indent~~and Facult\'e des Sciences,
     Univ. de l'Etat Mons, Av. Maistriau 19, B-7000 Mons, Belgium
    \label{AIM}}
\titlefoot{Physics Laboratory, University of Athens, Solonos Str.
     104, GR-10680 Athens, Greece
    \label{ATHENS}}
\titlefoot{Department of Physics, University of Bergen,
     All\'egaten 55, NO-5007 Bergen, Norway
    \label{BERGEN}}
\titlefoot{Dipartimento di Fisica, Universit\`a di Bologna and INFN,
     Via Irnerio 46, IT-40126 Bologna, Italy
    \label{BOLOGNA}}
\titlefoot{Centro Brasileiro de Pesquisas F\'{\i}sicas, rua Xavier Sigaud 150,
     BR-22290 Rio de Janeiro, Brazil \\
     \indent~~and Depto. de F\'{\i}sica, Pont. Univ. Cat\'olica,
     C.P. 38071 BR-22453 Rio de Janeiro, Brazil \\
     \indent~~and Inst. de F\'{\i}sica, Univ. Estadual do Rio de Janeiro,
     rua S\~{a}o Francisco Xavier 524, Rio de Janeiro, Brazil
    \label{BRASIL}}
\titlefoot{Comenius University, Faculty of Mathematics and Physics,
     Mlynska Dolina, SK-84215 Bratislava, Slovakia
    \label{BRATISLAVA}}
\titlefoot{Coll\`ege de France, Lab. de Physique Corpusculaire, IN2P3-CNRS,
     FR-75231 Paris Cedex 05, France
    \label{CDF}}
\titlefoot{CERN, CH-1211 Geneva 23, Switzerland
    \label{CERN}}
\titlefoot{Institut de Recherches Subatomiques, IN2P3 - CNRS/ULP - BP20,
     FR-67037 Strasbourg Cedex, France
    \label{CRN}}
\titlefoot{Now at DESY-Zeuthen, Platanenallee 6, D-15735 Zeuthen, Germany
    \label{DESY}}
\titlefoot{Institute of Nuclear Physics, N.C.S.R. Demokritos,
     P.O. Box 60228, GR-15310 Athens, Greece
    \label{DEMOKRITOS}}
\titlefoot{FZU, Inst. of Phys. of the C.A.S. High Energy Physics Division,
     Na Slovance 2, CZ-180 40, Praha 8, Czech Republic
    \label{FZU}}
\titlefoot{Dipartimento di Fisica, Universit\`a di Genova and INFN,
     Via Dodecaneso 33, IT-16146 Genova, Italy
    \label{GENOVA}}
\titlefoot{Institut des Sciences Nucl\'eaires, IN2P3-CNRS, Universit\'e
     de Grenoble 1, FR-38026 Grenoble Cedex, France
    \label{GRENOBLE}}
\titlefoot{Helsinki Institute of Physics, HIP,
     P.O. Box 9, FI-00014 Helsinki, Finland
    \label{HELSINKI}}
\titlefoot{Joint Institute for Nuclear Research, Dubna, Head Post
     Office, P.O. Box 79, RU-101 000 Moscow, Russian Federation
    \label{JINR}}
\titlefoot{Institut f\"ur Experimentelle Kernphysik,
     Universit\"at Karlsruhe, Postfach 6980, DE-76128 Karlsruhe,
     Germany
    \label{KARLSRUHE}}
\titlefoot{Institute of Nuclear Physics and University of Mining and Metalurgy,
     Ul. Kawiory 26a, PL-30055 Krakow, Poland
    \label{KRAKOW}}
\titlefoot{Universit\'e de Paris-Sud, Lab. de l'Acc\'el\'erateur
     Lin\'eaire, IN2P3-CNRS, B\^{a}t. 200, FR-91405 Orsay Cedex, France
    \label{LAL}}
\titlefoot{School of Physics and Chemistry, University of Lancaster,
     Lancaster LA1 4YB, UK
    \label{LANCASTER}}
\titlefoot{LIP, IST, FCUL - Av. Elias Garcia, 14-$1^{o}$,
     PT-1000 Lisboa Codex, Portugal
    \label{LIP}}
\titlefoot{Department of Physics, University of Liverpool, P.O.
     Box 147, Liverpool L69 3BX, UK
    \label{LIVERPOOL}}
\titlefoot{LPNHE, IN2P3-CNRS, Univ.~Paris VI et VII, Tour 33 (RdC),
     4 place Jussieu, FR-75252 Paris Cedex 05, France
    \label{LPNHE}}
\titlefoot{Department of Physics, University of Lund,
     S\"olvegatan 14, SE-223 63 Lund, Sweden
    \label{LUND}}
\titlefoot{Universit\'e Claude Bernard de Lyon, IPNL, IN2P3-CNRS,
     FR-69622 Villeurbanne Cedex, France
    \label{LYON}}
\titlefoot{Univ. d'Aix - Marseille II - CPP, IN2P3-CNRS,
     FR-13288 Marseille Cedex 09, France
    \label{MARSEILLE}}
\titlefoot{Dipartimento di Fisica, Universit\`a di Milano and INFN-MILANO,
     Via Celoria 16, IT-20133 Milan, Italy
    \label{MILANO}}
\titlefoot{Dipartimento di Fisica, Univ. di Milano-Bicocca and
     INFN-MILANO, Piazza delle Scienze 2, IT-20126 Milan, Italy
    \label{MILANO2}}
\titlefoot{Niels Bohr Institute, Blegdamsvej 17,
     DK-2100 Copenhagen {\O}, Denmark
    \label{NBI}}
\titlefoot{IPNP of MFF, Charles Univ., Areal MFF,
     V Holesovickach 2, CZ-180 00, Praha 8, Czech Republic
    \label{NC}}
\titlefoot{NIKHEF, Postbus 41882, NL-1009 DB
     Amsterdam, The Netherlands
    \label{NIKHEF}}
\titlefoot{National Technical University, Physics Department,
     Zografou Campus, GR-15773 Athens, Greece
    \label{NTU-ATHENS}}
\titlefoot{Physics Department, University of Oslo, Blindern,
     NO-1000 Oslo 3, Norway
    \label{OSLO}}
\titlefoot{Dpto. Fisica, Univ. Oviedo, Avda. Calvo Sotelo
     s/n, ES-33007 Oviedo, Spain
    \label{OVIEDO}}
\titlefoot{Department of Physics, University of Oxford,
     Keble Road, Oxford OX1 3RH, UK
    \label{OXFORD}}
\titlefoot{Dipartimento di Fisica, Universit\`a di Padova and
     INFN, Via Marzolo 8, IT-35131 Padua, Italy
    \label{PADOVA}}
\titlefoot{Rutherford Appleton Laboratory, Chilton, Didcot
     OX11 OQX, UK
    \label{RAL}}
\titlefoot{Dipartimento di Fisica, Universit\`a di Roma II and
     INFN, Tor Vergata, IT-00173 Rome, Italy
    \label{ROMA2}}
\titlefoot{Dipartimento di Fisica, Universit\`a di Roma III and
     INFN, Via della Vasca Navale 84, IT-00146 Rome, Italy
    \label{ROMA3}}
\titlefoot{DAPNIA/Service de Physique des Particules,
     CEA-Saclay, FR-91191 Gif-sur-Yvette Cedex, France
    \label{SACLAY}}
\titlefoot{Instituto de Fisica de Cantabria (CSIC-UC), Avda.
     los Castros s/n, ES-39006 Santander, Spain
    \label{SANTANDER}}
\titlefoot{Dipartimento di Fisica, Universit\`a degli Studi di Roma
     La Sapienza, Piazzale Aldo Moro 2, IT-00185 Rome, Italy
    \label{SAPIENZA}}
\titlefoot{Inst. for High Energy Physics, Serpukov
     P.O. Box 35, Protvino, (Moscow Region), Russian Federation
    \label{SERPUKHOV}}
\titlefoot{J. Stefan Institute, Jamova 39, SI-1000 Ljubljana, Slovenia
     and Laboratory for Astroparticle Physics,\\
     \indent~~Nova Gorica Polytechnic, Kostanjeviska 16a, SI-5000 Nova Gorica, Slovenia, \\
     \indent~~and Department of Physics, University of Ljubljana,
     SI-1000 Ljubljana, Slovenia
    \label{SLOVENIJA}}
\titlefoot{Fysikum, Stockholm University,
     Box 6730, SE-113 85 Stockholm, Sweden
    \label{STOCKHOLM}}
\titlefoot{Dipartimento di Fisica Sperimentale, Universit\`a di
     Torino and INFN, Via P. Giuria 1, IT-10125 Turin, Italy
    \label{TORINO}}
\titlefoot{Dipartimento di Fisica, Universit\`a di Trieste and
     INFN, Via A. Valerio 2, IT-34127 Trieste, Italy \\
     \indent~~and Istituto di Fisica, Universit\`a di Udine,
     IT-33100 Udine, Italy
    \label{TU}}
\titlefoot{Univ. Federal do Rio de Janeiro, C.P. 68528
     Cidade Univ., Ilha do Fund\~ao
     BR-21945-970 Rio de Janeiro, Brazil
    \label{UFRJ}}
\titlefoot{Department of Radiation Sciences, University of
     Uppsala, P.O. Box 535, SE-751 21 Uppsala, Sweden
    \label{UPPSALA}}
\titlefoot{IFIC, Valencia-CSIC, and D.F.A.M.N., U. de Valencia,
     Avda. Dr. Moliner 50, ES-46100 Burjassot (Valencia), Spain
    \label{VALENCIA}}
\titlefoot{Institut f\"ur Hochenergiephysik, \"Osterr. Akad.
     d. Wissensch., Nikolsdorfergasse 18, AT-1050 Vienna, Austria
    \label{VIENNA}}
\titlefoot{Inst. Nuclear Studies and University of Warsaw, Ul.
     Hoza 69, PL-00681 Warsaw, Poland
    \label{WARSZAWA}}
\titlefoot{Fachbereich Physik, University of Wuppertal, Postfach
     100 127, DE-42097 Wuppertal, Germany
    \label{WUPPERTAL}}
\addtolength{\textheight}{-10mm}
\addtolength{\footskip}{5mm}
\clearpage
\headsep 30.0pt
\end{titlepage}
%
\pagenumbering{arabic} 
\setcounter{footnote}{0} %
\large

\section{Introduction}

\subsection{Motivations}
The $R$--parity symmetry plays an essential role in the construction
of supersymmetric theories of interactions, such as the  Minimal 
Supersymmetric extension of the Standard Model (MSSM)~\cite{mssm}.
The conservation  of $R$--parity is closely
related to the conservation of lepton ($L$) and baryon ($B$) numbers
and the multiplicative quantum number associated to the $R$--parity
symmetry is defined by $R_p=(-1)^{3B+L+2S}$ for a particle with
spin~$S$~\cite{fayet}. Standard particles have even $R$--parity,
and the corresponding superpartners have odd $R$--parity.
The conservation of $R$--parity guarantees that 
the new spin--0 sfermions cannot
be directly exchanged between standard fermions. It implies 
that the new sparticles ($R_p=-1$) can only be pair-produced, 
and that the decay of a sparticle should lead to another one,
or an odd number of them. Then, it ensures the stability of
the Lightest Supersymmetric Particle (LSP). 
The MSSM is designed to conserve \mbox{$R$--parity:} it is phenomenologically
justified by proton decay constraints, and by the hope that a neutral
LSP will provide a good dark matter candidate. 

One of the major consequences  of the $R$--parity violation is 
obviously that the LSP is no longer stable since it 
is allowed to decay to standard fermions.
This fact modifies the signatures of the supersymmetric particle
production compared to the expected signatures in case of $R$--parity
conservation.
In any case, whether it turns out to be absolutely conserved or not,
$R$--parity plays an essential role in the study of the phenomenological
implications of supersymmetric theories.

In complementarity with the searches for supersymmetric particles
in the hypothesis of $R$--parity conservation,
direct searches for  \mbox{$R$--parity} \mbox{violation (\rpv)} signatures
in sparticle production have been performed by
the LEP2 experiments~\cite{paperlle,ALO}.
No evidence for supersymmetric particle production has been  observed
so far, independently of the hypothesis on  $R$--parity.
In 1998, the LEP centre-of-mass energy reached 189~GeV, and
an integrated luminosity of 158~pb$^{-1}$ was 
collected by the DELPHI experiment.
The results of the searches for pair production of supersymmetric 
particles under the hypothesis of \mbox{$R$--parity} violating
couplings between sleptons and leptons, performed with the data 
collected by DELPHI in 1997 at a centre-of-mass energy of
183~GeV~\cite{paperlle}, are updated by the analyses
of the data recorded in 1998 presented in this paper.

\subsection{\boldmath$R$--parity violation in the MSSM \label{couplings}}
\mbox{The \rpv} superpotential~\cite{weinberg} contains three trilinear terms, 
two violating $L$ conservation, and one violating $B$ conservation.  
We consider here only the  
$\lambda_{ijk}$\LiLjEk\ term ($i, j, k$ are generation indices,
$\rm L$ (${\bar{\rm E}}$) denote the lepton doublet (singlet) superfields) 
which couples the sleptons to
the leptons; since ${\lambda}_{ijk} = -{\lambda}_{jik}$,
there are nine independent ${\lambda}_{ijk}$ couplings. 
Upper limits on the $\lambda_{ijk}$ couplings can be derived from 
indirect searches of \mbox{$R$--parity} violating 
effects~\cite{barger89}--\cite{gdrpv}, 
assuming that only one $\lambda_{ijk}$ is
dominant at a time. For example, charged current universality allows
a limit on \Labb\ to be derived: 
\mbox{\Labb\ $< 0.049 \times {\rm m_{\tilde e_R} \over 100\ {GeV}/c^2}$}
and the upper limits on the neutrino mass are used
to derive a limit on \Lacc:
\mbox{\Lacc\ $< 0.006 \times \sqrt{\rm m_{\tilde \tau_R} \over
 100\ {GeV}/c^2}$}~\cite{dreiner99}.
Taking into account recent data on neutrino masses and mixings,
smaller values of the upper limits on several  $\lambda_{ijk}$
have been derived, all being over 0.0007 
(for m$_{\tilde \ell} = 100$~\GeVcc)~\cite{bhatta99}.
In the analyses described here, only one $\lambda_{ijk}$ was
assumed to be dominant and its upper bound has been taken into
account. 

The relevant MSSM parameters for \mbox{these  \rpv}  searches are:
M$_1$, M$_2$, the U(1) and SU(2) gaugino mass 
at the electroweak scale,
m$_0$, the scalar common mass at the GUT scale, 
$\mu$, the mixing mass term of the Higgs doublets at the electroweak
scale and \tanb, the ratio of the
vacuum expectation values of the two Higgs doublets. 
The unification of the gaugino masses at the GUT scale, 
which implies 
M$_1={ 5\over3}{\mathrm tan}^2\theta_W $M$_2
\simeq {1\over2}$M$_2$ at the electroweak scale,
is assumed in the study of production and/or
decay processes involving neutralinos and charginos.

We assume that the running of  \mbox{the \rpv} couplings from
the GUT scale to the electroweak does not have a significant
effect on the evolution of the gaugino and fermion masses. This is
an assumption that will be reconsidered once detailed theoretical
calculations on this subject  become available.

\subsection{\boldmath$R$--parity violating decays}
This paper presents the searches for pair produced
gauginos and sfermions.
In case of pair production $R_p$ is
conserved at the production vertex; the cross-sections do not
depend on the \rpv\ couplings. The $R$--parity violation
affects only the decay of sparticles.

Two types of supersymmetric particle decays are considered.
First, the {\it direct decay}, corresponding to the   
\mbox{sfermion \rpv} direct decay into two standard fermions,
or to the neutralino (chargino)
decay into a fermion and a virtual sfermion which then 
decays into two standard fermions.
Second, the {\it indirect decay} corresponding 
to the supersymmetric
particle cascade decay through $R$--parity conserving
vertices to  on-shell supersymmetric particles down to a 
lighter supersymmetric particle decaying via one \LLE\ coupling.

The direct  decay of a neutralino or a chargino 
via a dominant \Lijk\ coupling leads to 
purely leptonic decay products, with or without neutrinos
(\mbox{$\ell \ell^\prime \nu$}, $\ell \ell^{\prime} \ell^{\prime\prime}$, 
\mbox{$\ell \nu \nu$}). 
The indirect
decay of a heavier neutralino or a chargino adds jets and/or leptons to the 
leptons produced in the LSP decay.
 
The sneutrino direct decay gives two charged leptons: 
via \Lijk\, only the
\snu$_i$ and \snu$_j$ are allowed to decay directly to 
$\ell^{\pm}_{j}\ell^{\mp}_{k}$ and $\ell^{\pm}_{i}\ell^{\mp}_{k}$ respectively.
The charged slepton direct decay gives one neutrino and one
charged lepton (the lepton flavour may be different from the slepton one):
the supersymmetric partner of the right-handed lepton  $\slep_{kR}$ decays 
directly into  $\nu_{iL} \ell_{jL}$ or $\ell_{iL} \nu_{jL}$, and 
the supersymmetric partner of the left-handed lepton
$\slep_{i(j)L}$ decays into $\bar \nu_{j(i)L} \ell_{kR}$. 

The indirect decay of a sneutrino (resp. charged slepton)  into the
lightest neutralino
and  a neutrino (resp. charged lepton) leads to a purely leptonic
final state: 
two charged leptons and two neutrinos (resp. three charged leptons  
and a neutrino).
The indirect decay of a slepton into a chargino and its isospin
partner was not considered, and the direct decay of charged slepton
is not studied here.

When the charged leptons are $\tau$, additional
neutrinos are  generated in  the $\tau$ decay, producing more
missing energy in the decay and leading to a smaller number of
charged leptons in the final state.

\section{Data samples}
The total integrated luminosity collected by the DELPHI
detector~\cite{delphidet} during 1998 at centre-of-mass energies around
189~\GeV\ was 158~pb$^{-1}$.
An integrated luminosity of 153~pb$^{-1}$ has been analysed,
corresponding to high quality data, 
with the tracking detectors and the electromagnetic calorimeters
in good working condition.

To evaluate background contaminations,
different contributions coming from the \mbox{Standard} Model processes  
were considered. The Standard Model events were produced by  
the following generators:
\begin{malist}
\item {$\gamma\gamma$ events:}
{\tt BDK}~\cite{bdk} for $\gamma\gamma\to \ell^+\ell^-$ processes,
and {\tt TWOGAM}~\cite{twogam} 
for $\gamma\gamma\to$ hadron processes; 
biased samples containing events with a minimal
transverse energy of 4~\GeV\ were used;
\item {two-fermion processes:}
{\tt BABAMC}~\cite{baba} and {\tt BHWIDE}~\cite{bhwd} for Bhabha scattering,
{\tt KORALZ}~\cite{koralz} for e$^+$e$^-\to \mu^+\mu^-(\gamma )$ and
for e$^+$e$^-\to \tau^+\tau^-(\gamma )$ 
and {\tt PYTHIA}~\cite{pythia} for e$^+$e$^-\to \rm q \rm\bar q(\gamma)$ 
events;
\item {four-fermion processes:}
{\tt EXCALIBUR}~\cite{excal} for all types of 
four fermion processes: non resonant ($\rm{f\bar ff^\prime\bar f^\prime}$), 
single resonant 
(Z$\rm f \rm\bar f$, W$\rm f \rm\bar f^\prime$) and doubly resonant (ZZ, WW)
({\tt PYTHIA} was used also for cross-checks).\\
\end{malist}


Signal events were generated with
the {\tt SUSYGEN 2.20} program~\cite{susygen}
followed by the full DELPHI simulation and reconstruction program
({\tt DELSIM}). A faster simulation
({\tt SGV}\footnote{{\it Simulation \`a Grande Vitesse }
{\tt http://home.cern.ch/~berggren/sgv.html}}) was 
used to check that the efficiencies were stable  
at points without full simulation compared to their
values at the nearest points determined with the full simulation.
The $R$--parity violating couplings were set close to their experimental
upper limit derived from the \mbox{indirect \rpv} 
searches~(see section~\ref{couplings}).

The \XOI\ and \XPMI\ pair production 
was considered at several points in the MSSM parameter space,
in order to scan neutralino masses from 15 to 80~\GeVcc\ and
chargino masses from 45 to 95~\GeVcc. Moreover, for a given 
mass, several samples with different components and production
processes were simulated.
The pair production of 
heavier neutralinos and charginos  has been taken
into account since one can profit from the threefold increase 
in luminosity compared to the 1997 data. 

For the study of slepton pair production, 
samples with sneutrino direct
decay and samples with sneutrino or charged slepton indirect decay
were generated for tan$\beta$ fixed at 1.5.
A \snu\ (\slep) mass range from 50 to 90~\GeVcc\ was
covered; in the case of indirect decay,
several ranges of mass difference between sleptons
and neutralinos were considered.

\section{Analysis descriptions}

\subsection{Analysis strategy and validity}
Any of the possible \mbox{\rpv} signals produced via 
one of the \Lijk\LiLjEk\ couplings can be
explored by the analyses described in this paper.
In the analyses performed considering a dominant \Lacc\ coupling, the
efficiencies and the rejection power are low,
due to the presence of several taus in the final state. 
The highest efficiencies and background reduction
are obtained if  \Labb\ is the dominant coupling.
For final states produced by  other $\lambda_{ijk}$, the detection
efficiencies lay between these two limiting cases.
Analyses are then performed considering both the  \Labb\ and the  \Lacc\  
couplings. 
The weakest limits were derived considering the results of the
analyses  performed assuming a dominant \Lacc\  coupling.
%
The studied final states are summarized in Table~\ref{cb.tab1}.\\
\begin{table}
\begin{center}
\begin{tabular}{|lccc|} \hline
processes & 
final states with \Labb & ~~ &
final states with \Lacc \\ \hline
\XOi\XOj, \XPk\XMl (direct    
& e$\mu$e$\mu$, e$\mu\mu\mu  ,\ \mu\mu\mu\mu$ + E$_{\rm miss}$ 
&~~
& e$\tau$e$\tau $, e$\tau\tau\tau  ,\ \tau\tau\tau\tau$ + E$_{\rm miss}$ \\
and indirect decays)    
& ($+ n\ell$) ($+m$ qq')        
&~~
& ($+ n\ell$) ($+m$ qq')         \\ \hline
\snue\asnue\  (direct decay) & $\mu\mu\mu\mu$ &~~& $\tau\tau\tau\tau$ \\       
\snut\asnut\  (direct decay) & ee$\mu\mu$     &~~& ee$\tau\tau$ \\ \hline
\snu\asnu\ (indirect decay) 
& e$\mu$e$\mu$, e$\mu\mu\mu ,\ \mu\mu\mu\mu$ + E$_{\rm miss}$ 
&~~
& e$\tau$e$\tau$, e$\tau\tau\tau ,\ \tau\tau\tau\tau$ + E$_{\rm miss}$\\ \hline
\slep$_R^+$\slep$_R^-$ (indirect decay) 
& e$\mu$e$\mu$, e$\mu\mu\mu  ,\ \mu\mu\mu\mu$ 
&~~
& e$\tau$e$\tau$, e$\tau\tau\tau ,\ \tau\tau\tau\tau$ \\
& + E$_{\rm miss}$ + $\ell^+ \ell^-$
&~~
& + E$_{\rm miss}$ + $\ell^+\ell^-$ \\ \hline
\end{tabular}
\caption{Pair production final states with \Labb\ or \Lacc.}
\label{cb.tab1}
\end{center}
\end{table}

It was supposed that the
Lightest Supersymmetric Particle (LSP) decays
within a few centimeters of the production vertex.
Since the mean LSP decay length depends on 
${\rm m}_{\chi}^{-5}$
(if the LSP is a gaugino), and on $\lambda_{ijk}^{-2}$,
this assumption has two consequences on the 
analyses described here.
First, they were not sensitive to light  
$\widetilde{\chi}$ (M$_{\tilde\chi_{LSP}} \leq\ $10~GeV/$c^2$).
Second,  analyses looking for neutralino decay products 
had a lower limit in the sensitivity of the
$\lambda$ coupling  of the order of $10^{-4}$; below this
value, in some area in the MSSM space, the lightest neutralino
has a non-negligible lifetime, and the corresponding 
event topology was not selected by the analyses.  
Inside the  validity domain defined by the upper bound from 
indirect searches \mbox{of \rpv} effects 
and the lower bound due to the LSP flight, 
the coupling value has no influence on the  efficiency of the analyses.

\subsection{General analysis description}
The applied  selections were based on the criteria presented 
in~\cite{paperlle},
using mainly topological criteria, missing quantities, lepton identification
and kinematic properties, and jet characteristics. 
Compared to the previous analyses, the electron identification
has been improved
at high energies and  
in the forward regions of the detector.

As already mentioned, indirect decays of gaugino pairs
can add two or more jets to the leptons and
missing energy final state, 
from the hadronic decay of W$^*$ and Z$^*$. 
Moreover, in the case of the \Lacc\ coupling,
thin jets are produced in
$\tau$ decay. The jets were reconstructed
with the {\tt DURHAM}~\cite{Durham} algorithm. 
In order to cover the different topologies, the jet number was not
fixed,
and the jet charged multiplicity could be low (for instance thin
jets with one charged
particle), or could be zero in case of neutral jets.
In the following, the transition value
of the y$_{cut}$  in the {\tt DURHAM} algorithm
at which  the event changes from a $n$-jet
to a $(n-1)$-jet configuration is noted y$_{(n-1)n}$.

After a brief description of the \Labb\ analyses, 
the selection procedures when \Lacc\ is the dominant coupling
constant are detailed in the following sections.

\subsection{Analyses applied in case of \Labb\ coupling}
As already mentioned, these analyses 
were based on the  selection procedure described 
in~\cite{paperlle}; they are not deeply detailed here.

\subsubsection{Gaugino and slepton indirect decay searches}
One analysis was designed to select leptonic
channels with missing energy, 
with or without jets,
in order to study gaugino decays and slepton indirect decays.
Events with charged multiplicity greater than
three and at least two charged particles with a
polar angle between 40$^{\circ}$ and 
140$^{\circ}$ were selected. The
missing transverse momentum, \Ptm, had to be greater than  5 \GeVc\ and 
the polar angle of the missing momentum to be between 20\mydeg\ and
160\mydeg. The missing energy had to be at least 0.2$\sqrt{s}$.
This set of criteria reduced mainly the background coming from Bhabha
scattering and two-photon processes.
Then, requirements based on the lepton characteristics were applied:
\begin{malist}
\item at least two identified muons were required;
\item the energy of the most energetic identified 
lepton had to be greater than 0.1$\sqrt{s}$;
\item an isolation criterion was imposed 
for the identified leptons (no other charged particle in 
 a half cone of seven degrees around the lepton);
\item at least two of the identified leptons had to be leading particles
in the jets.
\end{malist}
One event remained in the data, compared to 1.1$\pm$0.3 expected  
from Standard Model processes contributing to the background 
(0.6, 0.3 and 0.2 from four-fermion, $\mu^+\mu^- (\gamma)$ and 
$\gamma\gamma \rightarrow \mu^+\mu^-$ processes  respectively).
The selection efficiencies were in the range 35--60\% for 
gaugino decays, in the range  50--60\% for 
sneutrino indirect decays, and in the range 20--50\%
for charged slepton indirect decays.

\subsubsection{Sneutrino direct decay search}
A different selection was used  to search for final states 
with no missing energy and at least two muons, resulting
from sneutrino direct decays  via \Labb .
The thrust value had to be less than 0.95 and 
the polar angle of the thrust axis had to be between 25\mydeg and 155\mydeg.
The total energy from charged particles  had 
to be greater than 0.33$\sqrt{s}$, the
missing transverse momentum had to be greater than  2~\GeVc\
and the missing energy to be  less than  0.55$\sqrt{s}$.
The charged multiplicity had to be four or six with
the total event charge equal to 0. At least two muons were required 
and no other charged particle in a half cone of 20\mydeg\ around 
each lepton was demanded.
One event remained in the data after these criteria with 2.7$\pm$0.4 
expected from
standard background processes, mainly from the 
$\ell^+\ell^-\ell^{'+}\ell^{'-}$ final states (1.4$\pm$0.2), 
and from the  $\gamma\gamma \rightarrow \mu^+\mu^-$ 
process (1.2$\pm$0.4). The efficiencies were from 62\% to 51\% in the
explored sneutrino mass range of 60--90~\GeVcc.

\subsection{Analyses applied in case of \Lacc\ coupling}

\subsubsection{Preselection criteria for \Lacc\ analyses\label{sec:pairpresel}}

\begin{figure}[h]
\begin{center}
\epsfig{file=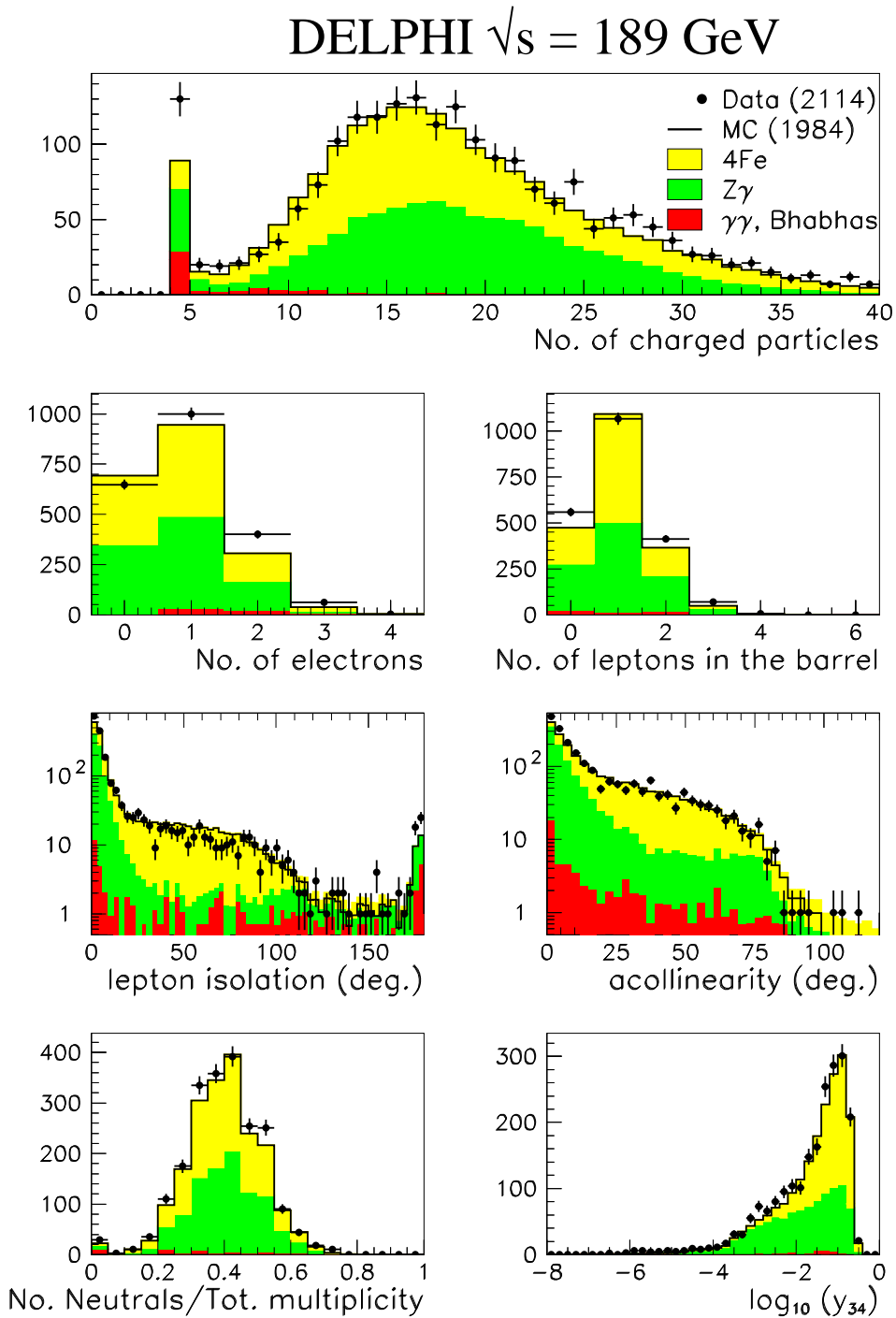,width=14cm} 
\caption{Distributions, after the preselection applied for
the \Lacc\ analyses,
of the number of charged particles, the number
of well identified electrons in the event, the
number of identified leptons with a polar angle
between 40$^\circ$ and 140$^\circ$, the lepton
isolation angle, the acollinearity, the ratio of the number
of neutral particles to the total event multiplicity,
and the log$_{10}(\rm y_{34})$. The  
 black dots show the real data distributions, and the shaded histograms
 the expected background from Standard Model processes.}  
\label{distrib}
\end{center}
\vspace{-0.5cm}
\end{figure}

In the  search for pair production of gauginos and sleptons
in case of a dominant \Lacc\ coupling, the following criteria
were required:
\begin{malist}
\item at least one identified lepton;
\item more than three charged particles
and at least two of them with a polar angle between 40$^{\circ}$
and 140$^{\circ}$;
\item the  total energy and the energy from charged particles 
greater than 0.18$\sqrt{s}$ and 0.16$\sqrt{s}$ respectively;
\item the missing \Pt\ greater than  5 \GeVc;  
\item the polar angle of the missing momentum  
between 27\mydeg\ and 153\mydeg.
\end{malist}
This was efficient in suppressing the background coming from
Bhabha scattering  and  two-photon processes and in removing a large part 
of the \Zg\ contribution. After this preselection stage, 2114 events
were selected compared with 1984$\pm$11 expected from the background 
sources (see Figure~\ref{distrib}).
There was an excess of data mostly concentrated in the low charged 
multiplicity events where $\gamma\gamma$ events contributed
to the Standard Model background. 
A good agreement between data and the expected background was obtained 
when the contribution of   $\gamma\gamma$ events was further reduced
(see below).

\subsubsection{Neutralino and chargino search\label{sec:pairbosino}}
\begin{table}[bht]
  \begin{center}
    \begin{tabular}{|lll|c|c|} \hline
\multicolumn{3}{|c|}{{\bf Selection criteria}} 
               &  & \\ \cline{1-3} 
   &  &  & Data & MC \\  
   & 4$\leq $N$_{\rm charged} \leq 6$ & N$_{\rm charged}\geq 7 $& &  \\ \hline\cline{1-5}
acollinearity &  & $> 7^\circ$ & 1342 & 1301$\pm$8 \\ \hline
E$_{\rm cone}^{30^\circ}$ & $ \leq$ 0.5 E$_{\rm total}$ &$ \leq$ 0.4 E$_{\rm total}$
& 1146 & 1121$\pm$7  \\ \hline
N$_{\rm lepton}$ in  & & & & \\
the barrel & $\geq 1$ & $\geq 1 $ & 929 & 915$\pm$6 \\ \hline
E$^l_{\rm max} $& [2 GeV, 70 GeV]  & [5 GeV, 60 GeV] 
&652  & 665$\pm$5    \\  \hline
isolation &$ \Theta ^{\rm min}_{\ell \rm -charged\ particle} \geq 20^\circ$ 
          &$ \Theta ^{\rm max}_{\ell \rm -charged\ particle} \geq 6^\circ$ & & \\ 
          & if N$_{\rm charged}=$4 &  & &  \\ 
          &$ \Theta ^{\rm min}_{\ell \rm -charged\ particle} \geq 6^\circ$ 
          & $ \Theta ^{\rm max-1}_{\ell \rm -charged\ particle} \geq 10^\circ$ &   &  \\ 
         &  if N$_{\rm charged}=5, 6 $ & if N$_{\rm lepton}\geq 2$ 
&  &     \\ \cline{1-3}
N$_{\rm neutral} \leq $ & 10 & 15 &  &  \\ \cline{1-3}
N$_{\rm electron}$  & & $\geq 1 $ & 131 & 147$\pm$3 \\ \hline
E$_{\rm miss}$ &  $>$0.3$\sqrt{s}$ &$ >$ 0.3$\sqrt{s}$ & 96   &
               101$\pm$2 \\ \hline
$\rm log_{10}(\rm y_{23})$ & $\geq -2.7$ & $\geq -1.8 $ & &  \\
$\rm log_{10}(\rm y_{34})$ & $ \geq -4 $ & $\geq -2.3 $ & &  \\
$\rm log_{10}(\rm y_{45})$ &            & $\geq -3 $ &
16 & 14.7$\pm$0.7 \\ \hline
4 jets & & & & \\
E$^j_{\rm min} \times \theta^{j1,j2}_{\rm min}$ & $  \geq 1$ GeV.rad &  $
\geq 5$ GeV.rad &15 &13.9$\pm$0.6 \\ \hline
& & at least 1 jet with & &  \\
& &  1 or 2 charged &  &  \\ 
& &  particle(s)&  &  \\ 
&4 charged jets  & 4 charged jets if 4j  & &  \\
& if 4j or 5j & 4 or 5 charged  & & \\
&             & jets if 5j & 11 & 10.5$\pm$0.5  \\ \hline
\end{tabular}
\caption{Selection criteria used in the search for neutralino and
chargino decay
via \Lacc. $n$j means $n$-jet topology, and a charged jet means
a jet with at least one charged particle. 
The number of remaining data and Standard Model background events 
are reported; the quoted errors are statistical.}
    \label{cb-bosino-cut}
  \end{center}
\end{table}
Compared to the selection applied to 1997 data~\cite{paperlle},
it has been necessary to modify some criteria
and to distinguish between low and high multiplicity cases
in order to reach a higher purity.
For events with a charged particle multiplicity from four to six
(which corresponds to neutralino or chargino direct decay), the
following criteria were applied:
\begin{malist}
\item the energy in a cone of 30$^\circ$ around the beam axis was
restricted to be less than 50\% of the total visible energy;
\item the energy of the most energetic lepton had to be between 
2 and 70~GeV;
\item  there should be no other charged particle 
in a 10\mydeg\ (6\mydeg) half cone around any
identified lepton for a charged particle multiplicity equal to
four (five or six);
\item the number of neutral particles had to be less than or equal
to 10.
\end{malist}
For events  with a charged particle multiplicity greater than six
(corresponding to neutralino and chargino indirect decays),
the criteria were:
\begin{malist}
\item the acollinearity~\footnote{the acollinearity is computed 
between the two vectors corresponding to the sum of the particle
momenta  in each event hemisphere.} had to be greater than 7$^\circ$;
\item the energy in a cone of 30$^\circ$ around the beam axis was
restricted to be less than 40\% of the total visible energy;
\item the energy of the most energetic lepton had to be between 
5 and 60~GeV;
\item if there was only one identified lepton, 
no other charged particle  in a 6\mydeg\ half
cone around it was allowed; and
if there were more, 
there should not  be any other charged particle in a 10\mydeg\ half
cone around at least two of them;  
\item at least one well identified electron;
\item the number of neutral particles had to be less than or equal to
15.
\end{malist}
In both cases the missing energy had to be at least 30\% of the
available energy, and
the polar angle of at least one lepton  
had to be between 40$^\circ$  and 140$^\circ$.  
These criteria removed \Zg\ and hadronic \ZZ\ and 
\WW\ events.

The selection based on the jet characteristics and topologies was then
applied.
First, constraints have been imposed 
to y$_{(n-1)n}$ values to reduce, in particular the \Zg\ contribution.  
In events with more than six charged particles, at least one
jet with low charged particle multiplicity was required.
In four- or five-jet configurations, a minimum of four 
charged jets was required. In case of a four-jet topology,
a cut was applied on the value of 
E$^j_{{\rm min}} \times \theta^{j_aj_b}_{{\rm min}}$
where E$^j_{{\rm min}}$ is the energy of the least energetic jet, and
$\theta^{j_aj_b}_{{\rm min}}$ is the minimum angle between any pair of jets.
These requirements significantly reduced the  background from 
\Zg\ and \WW\ production.
The number of remaining real data
and background  events during the selection
are reported in Table~\ref{cb-bosino-cut}, and the contributions of the 
relevant Standard Model processes are detailed in Table~\ref{cb-bosino-sel}.
The main contribution comes from the \WW\ production, with a
semi-leptonic decay of the W pair, due to the
specific design of the analysis to be efficient for channels
with leptons (mainly taus) and jets in final states.
\begin{table}[hbt]
\begin{center}
\begin{tabular}{|l|c|c|c|c|c|c|c|c|}\hline
case & Data  & total MC    & $\rm q \rm\bar{q} (\gamma) $& $\tau^+\tau^- (\gamma)$ & 
  \Zee & \Wev &  \WW & \ZZ \\ \hline
Low &2     & 1.8$\pm$0.2 &  0. &  0.12$\pm$0.12 & 0.42$\pm$0.14 &
  0.            &  0.77$\pm$0.14 & 0.41$\pm$0.08 \\ \hline
High & 9     & 8.7$\pm$0.5  & 0.14$\pm$0.09 & 0.  &  0.06$\pm$0.06 &  
  0.05$\pm$0.02 &  8.27$\pm$0.44 & 0.21$\pm$0.07 \\ \hline
 \end{tabular}
\caption{Standard Model background contributions to the neutralino and
  chargino pair production analysis (\Lacc). The results in the row labelled
  ``Low'' (``High'') are obtained with the selection applied to the low (high)
  multiplicity events. The quoted errors are statistical.}
\label{cb-bosino-sel}
\end{center}
\end{table}

Using the events produced with {\tt DELSIM}, selection efficiencies
have been studied on \XOI\XOI\ and \XPI\XMI\ signals. In order
to benefit from the high luminosity, all 
\mbox{$\rm e^+ \rm e^-$ \Ra \XOi\XOj },
\mbox{$\rm e^+ \rm e^-$ \Ra \XPk\XMl }
processes which contribute
significantly have been simulated, at each \mbox{MSSM} point of this
study. {\tt SUSYGEN} followed by {\tt SGV} was used for the scan. Then
a global event selection efficiency was determined for each
point,  since the 
performed analyses were sensitive to many different topologies. 
The global selection efficiencies obtained with {\tt SGV} simulated events
have been cross-checked at several points 
with {\tt DELSIM} simulated events. The efficiencies laid between
 18\% and 40\%.

\subsubsection{Sneutrino and charged slepton searches}
Considering the \Lacc\ coupling, 
searches for sneutrino pair production  and subsequent  
direct (\SNU \Ra $\ell^+ \ell^-$) or indirect  (\SNU \Ra
\XOI$\nu$) decay and searches for
charged slepton pair production decaying  indirectly 
(\slep \Ra \XOI$\ell$) have been performed. 
In these different searches, a large amount of energy
is missing in the final states, due to neutrinos (from $\tau$
and/or \XOI\ decays), except in the case of \snut\asnut\ direct decay
search (ee$\tau\tau$ final state).
Two different analyses were then performed, one applied to the
channels with a large amount of missing energy, and the other one
dedicated to the ee$\tau\tau$ channel, with less missing energy.\\

\noindent{\bf $\bullet$ Analysis for channels with high value of missing energy}\\
The selection procedure was close to the one
applied to low charged particle multiplicity events in the search for
neutralino and chargino pair production. The same event
charac\-teristics
were used.
A large amount of missing energy was required, but 
only events with four to eight charged 
particles were selected.
The criteria are listed in Table~\ref{cb-snu1-cut};
the number of observed  events
and expected ones from the Standard Model background
during the selection procedure is also given.
At the end, one event remains in the data compared
to 2.1$\pm$0.3 from the SM processes. The
relevant contributions are listed in Table~\ref{cb-snu1-sel}. 
The four-fermion contributions have been checked also with
{\tt EXCALIBUR}, and apart from the \mbox{WW-like}
processes, the two other important background sources were the
ee$\tau\tau$ and ee$\mu\mu$ final states.
\begin{table}[h]
\begin{center}
\begin{tabular}{|l|c|c|}\hline \hline
\multicolumn{1}{|c|}{\bf Selection criteria} & Data & MC  \\ \hline
N$_{\rm charged} \leq $8  & & \\ 
E$_{\rm miss} >$30\%$\sqrt{s}$ & 120 & 106.6$\pm$3.4  \\ \hline
$2 \leq $ E$^{\ell}_{\rm max} \leq 70$ GeV     & 88 & 89.2$\pm$2.9 \\ \hline
$ \Theta ^{\rm min}_{\ell \rm -charged\ particle} \geq 20^\circ$ if N$_{\rm
  charged}=$4 & &  \\ 
$ \Theta ^{\rm min}_{\ell \rm -charged\ particle} \geq 6^\circ$ if N$_{\rm charged}>$4 & 62 & 61.5$\pm$2.4 \\ \hline
N$_{\rm neutral} \leq $10 & 55 &  52.3$\pm$2.2  \\ \hline
at least 1 lepton in the barrel & 25 &  22.4$\pm$1.3 \\ \hline
log$_{10}(\rm y_{23}) \geq -2.7$ &  &  \\
log$_{10}(\rm y_{34}) \geq -4 $ & 5 & 4.4$\pm$0.4 \\ \hline
in 4-jet events: & &   \\
$\theta^{j1,j2}_{\rm min} \geq 20^\circ $ &  &  \\
at least 1 jet with 1 or 2 charged particles  & 1 & 2.1$\pm$0.3 \\ \hline
\end{tabular}
\caption{Selection criteria used in the search for slepton pair
production  \mbox{with \rpv} decay
via \Lacc. The number of remaining data and Standard Model background events 
are reported; the quoted errors are statistical.}
\label{cb-snu1-cut}
\end{center}
\end{table}

\begin{table}[tbh]
\begin{center}
\begin{tabular}{|c|c|c|c|c|c|}\hline
 Data  & total MC    & $\tau^+\tau^- (\gamma)$ & \Zee &  \WW & \ZZ \\ \hline
 1    & 2.13$\pm$0.27  & 0.12$\pm$0.12 &
 0.54$\pm$0.16 &  1.13$\pm$0.16 & 0.34$\pm$0.08 \\ \hline
 \end{tabular}
\caption{Standard Model background contribution to the slepton pair
 production analysis (\Lacc); the quoted errors are statistical.}
\label{cb-snu1-sel}
\end{center}
\end{table}
  
For the 4$\tau$ channel produced in \snue\asnue\ decay, the
efficiencies
were between
27\% and 31\%. The sneutrino indirect decay efficiencies ranged
from 17\% (m$_{\tilde\nu} =$~50~\GeVcc, 
m$_{\widetilde{\chi}^0} =$~23~\GeVcc) 
to  36\%  (m$_{\tilde\nu} =$~80~\GeVcc, 
m$_{\widetilde{\chi}^0} =$~60~\GeVcc).
The charged slepton indirect decay efficiencies wrre higher, due to
the presence of two additional charged leptons in the final state,
and laid between 33\% and 40\%.\\

\noindent{\bf $\bullet$ Analysis for channels with low value of missing energy}\\
In order to obtain higher efficiencies for the ee$\tau\tau$ channel,
the selection criteria were modified. In particular, 
the missing energy cut was reduced to 8\% of the available energy.
The number of charged particles
was restricted to be between four and six. Moreover, the lower
limit on the energy of the most energetic lepton was increased
to 20 GeV, and the isolation angle had to be greater than
10$^\circ$. Additional criteria
were used: 
the acollinearity had to be greater than 2$^\circ$, 
and  the presence of 
at least one  identified electron was required.
After the event selection 3 events remained, while 
2.3$\pm$0.3 events were expected from SM processes.
The main sources of background were the ee$\tau\tau$ (57\%)
and the ee$\mu\mu$ (29\%) four-fermion processes.
The efficiencies were between 38\% and 46\%.

\section{Interpretation of the results}
The results of the searches presented in this paper,
summarised in Table~\ref{tabres}, were
in agreement with the Standard Model expectation. 
They were used to
extend the previously excluded part of the  MSSM parameter space
and to update limits obtained with the analysis of the 1997 data
collected in DELPHI. 
\begin{table}[h]
\begin{center}
\begin{tabular}{|c|l|c|c|c|}\hline
Coupling & Process & Efficiency     & \multicolumn{2}{|c|}{Selected events} \\
         &         & range in \%    & Data & MC \\ \hline \hline
\Labb & \XOi\XOj, \XPk\XMl\ direct & &   &     \\
      & and indirect decays  & 35--60 &     &    \\
      & \snu\asnu\ indirect decay & 50--60 &  1  & 1.1$\pm$0.3  \\
      & \slep$^+$\slep$^-$ indirect decay & 20--50 & & \\ \cline{2-5}
      & \snum\asnum\ direct decay & 51--62 & 1 & 2.7$\pm$0.4 \\ \hline
\Lacc & \XOi\XOj\ direct decay  & 18--40 & 2 & 1.8$\pm$0.2  \\
      & \XPk\XMl\ indirect decay& 18--40 & 9 & 8.7$\pm$0.5  \\
         \cline{2-5}
      & \snue\asnue\ direct decay& 27--31 & & \\
      & \snu\asnu\ indirect decay& 17--36 & 1 & 2.1$\pm$0.3 \\
      & \slep$^+$\slep$^-$ indirect decay & 33--40 & & \\  \cline{2-5}
      & \snut\asnut\ direct decay& 38--46 & 3 & 2.3$\pm$0.3 \\ \hline
\end{tabular}
\caption{\LLE\ analyses: efficiency ranges in the different cases
studied, and data and Monte Carlo events remaining after the
applied selection.} 
\label{tabres}
\end{center}
\end{table}
In all the pair production processes studied, the weakest 
limits were derived from the results of
the \Lacc\ analyses, and are hence valid for any choice of 
dominant \Lijk\ coupling, provided
that the  coupling is strong enough for the LSP to decay within a few
 centimetres. \\

\begin{figure}[h]
\begin{center}
\epsfig{file=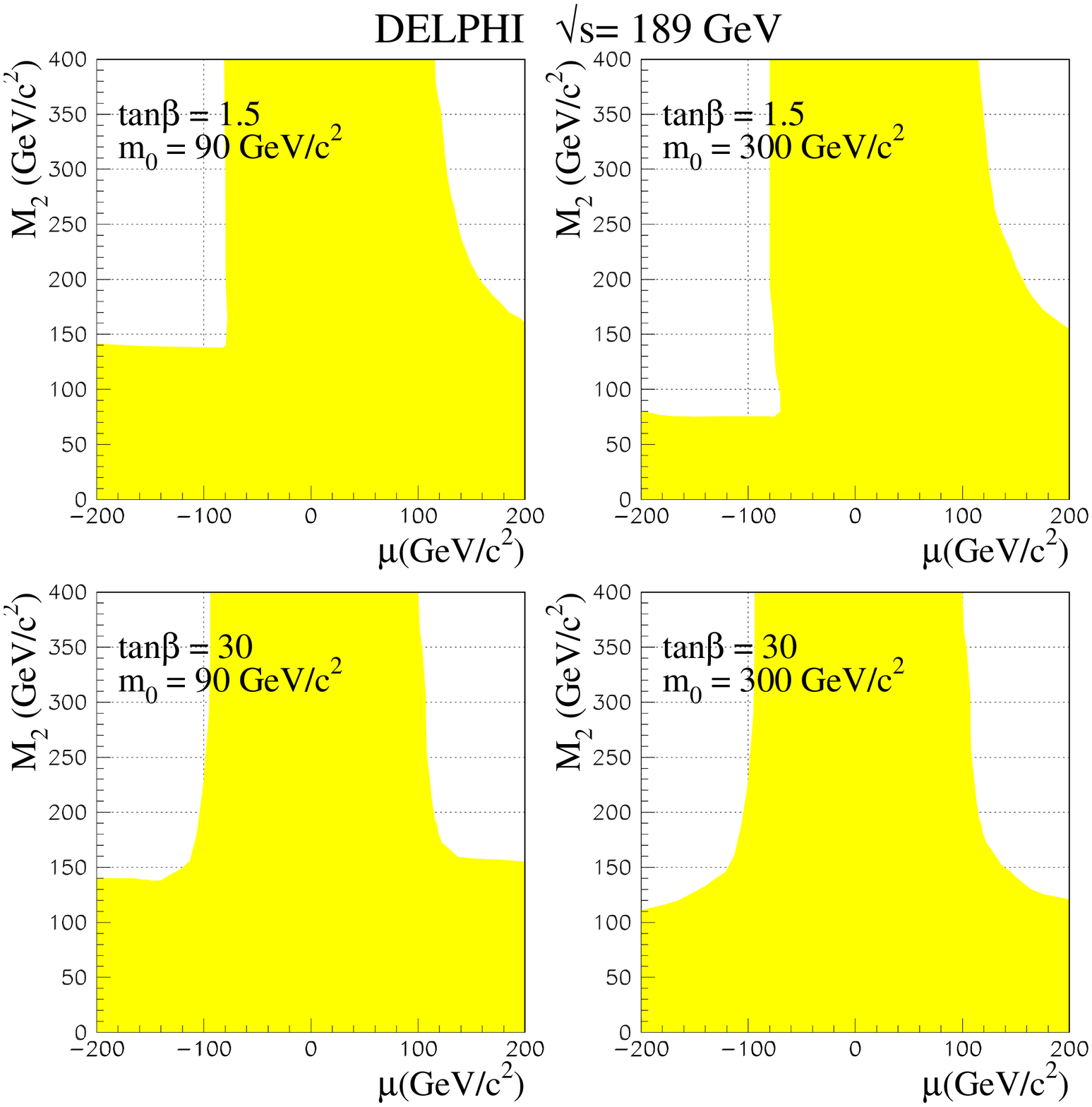,width=13cm}
\caption{
Decays through the $\lambda$ $R$--parity violating operator:
excluded regions at 95\%~C.L. in the $\mu$, M$_2$ parameter space
by the  neutralino and chargino searches  in DELPHI at 189~\GeV\
for two values of \tanb\ and two values of m$_{0}$.} 
\label{exclu189}

\epsfig{file=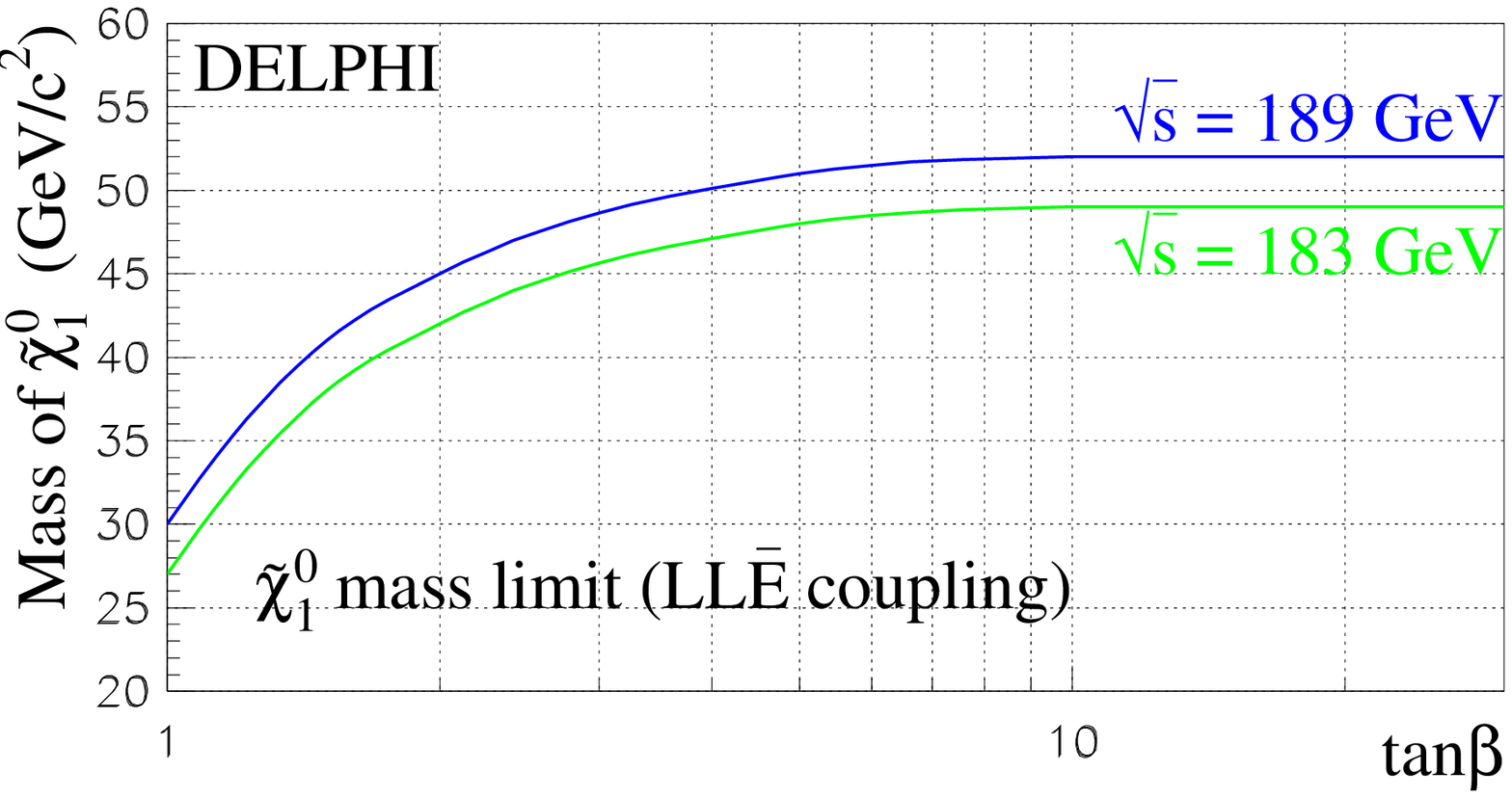,width=13cm}
\caption{The 
excluded lightest neutralino mass as a function of
\tanb\ at 95\% confidence level.
This limit is valid for all generation indices
$i$,$j$,$k$ of the $\lambda_{ijk}$ coupling and  
all values of m$_0$}
\label{chi0lim189}
\end{center}
\end{figure}

In the searches for neutralino and chargino
 pair production, the number of expected
events at each point of the explored MSSM parameter space was
obtained by:
\begin{center}  
N$_{\rm exp}=$ \lum $\times {\large \boldmath \epsilon_g} \times$ 
$\{ \sum_{i,j=1}^4 \sigma( \rm e^+ \rm e^-$ \Ra \XOi\XOj 
$) + \sum_{k,l=1}^2 \sigma( \rm e^+ \rm e^-$ \Ra \XPk\XMl $)  \} $
\end{center}
where \lum\ is the integrated luminosity, and 
$\large \boldmath \epsilon_g$ is the
global efficiency determined as explained in section~\ref{sec:pairbosino}.
This number has been compared to the number of signal
events, N$_{95}$, expected at a confidence level of 95\% in presence
of background~\cite{pdg}. 
All points which satisfied N$_{\rm exp} > $N$_{95}$
were excluded at 95\% C.L.
The excluded area
in $\mu$, M$_2$ planes obtained with the present searches
are shown in Fig.~\ref{exclu189}, for
m$_0$~=~90~\GeVcc\ (the t--channel contribution to the gaugino cross-sections
has an important effect),
m$_0$~=~300~\GeVcc\ (the t-channel
contribution vanishes) and tan$\beta$~=1.5, 30.
The smaller excluded area in the  $\mu$, M$_2$ planes for
a given tan$\beta$ is obtained for high  m$_0$ values. 

For each tan$\beta$, the highest value of neutralino mass
which can be excluded has been determined in the 
$\mu$, M$_2$ plane (--200~\GeVcc~$\leq \mu \leq$~200~\GeVcc, 
5~$<$ M$_2 \leq$~400~\GeVcc)
for several m$_0$~values varying up to 500~\GeVcc. 
The smaller excluded area in the  $\mu$, M$_2$ plane is obtained
for m$_0$~=~500~\GeVcc. The most
conservative mass limit was obtained for high 
m$_0$~values, for which it reaches a plateau.
The corresponding limit on neutralino mass as a function of tan$\beta$
is plotted in Fig.~\ref{chi0lim189}.
From these studies, a neutralino lighter than 30~\GeVcc\ was excluded at 
95\%~C.L. for $1 \leq $tan$\beta \leq 30$.
The same procedure was applied to determine the 
most conservative lower limit on the chargino masses. The result is less 
dependent on tan$\beta$, almost reaching the kinematic limit
for any value of tan$\beta$:
a chargino lighter than 94 GeV/$c^2$ was excluded at 95\% C.L.
Finally, using the same method,
a lower limit of 50~\GeVcc\ for the \XOII\ mass 
has been derived  at 95\% C.L. \\

\begin{figure}[h]
\begin{center}
\epsfig{file=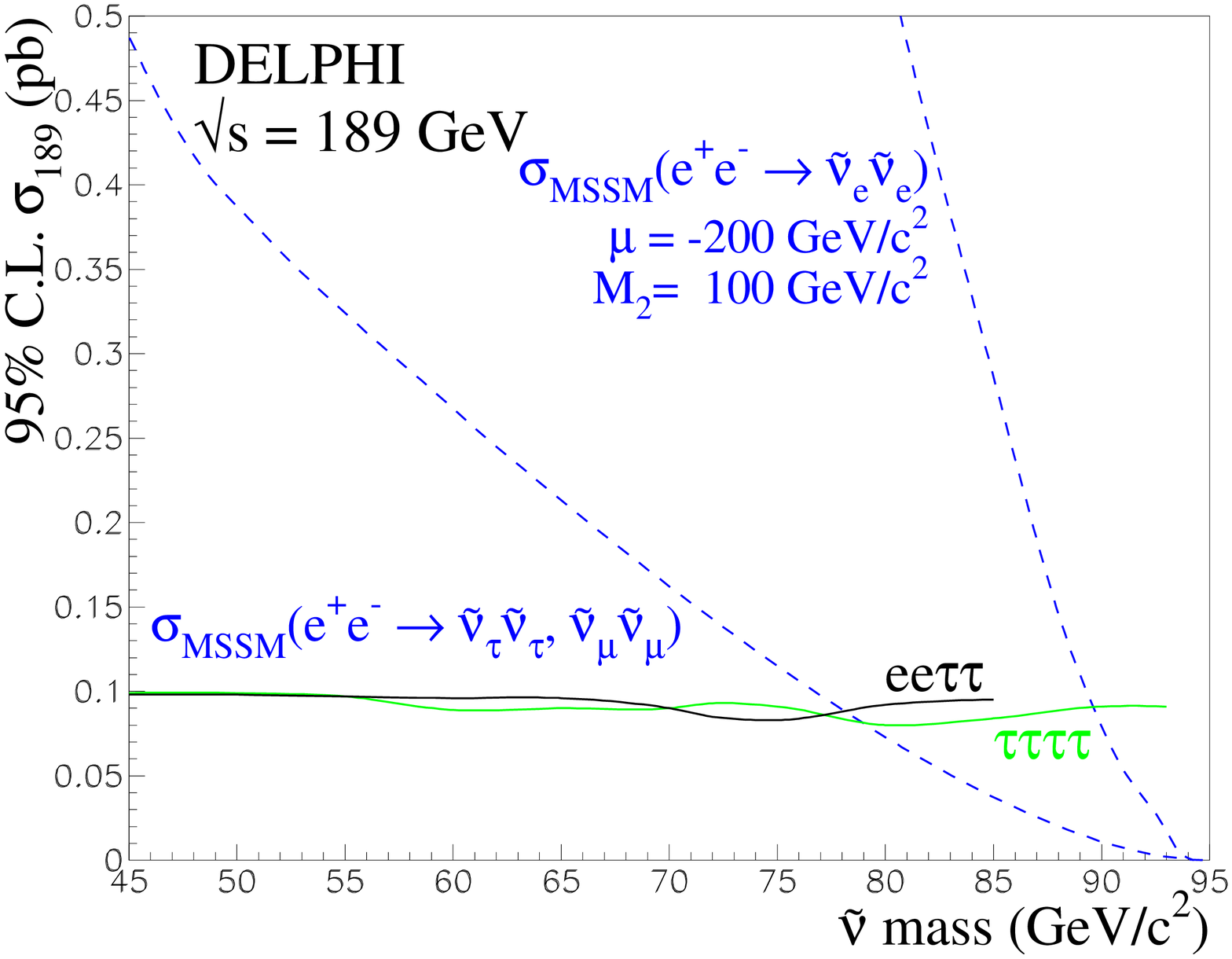,width=10cm}
\caption{Sneutrino direct decay with \Lacc\ coupling: limit on the
\snu\asnu\ production cross-section as a function of the mass
for two different final states. The MSSM cross-sections 
are  reported in order
to derive a limit on the sneutrino mass in the case of
\mbox{direct \rpv} decay.
The dashed lower curve corresponds to both  \snum\asnum\ and  
\snut\asnut\ cross-sections which depend only on the \snu\ mass. 
The dashed upper curve is the \snue\asnue\ 
cross-section obtained for $\mu = -200$~\GeVcc\ and $M_2 = 100$~\GeVcc,
the corresponding chargino mass lies between 90~and 120~\GeVcc.}
\label{snudir}

\epsfig{file=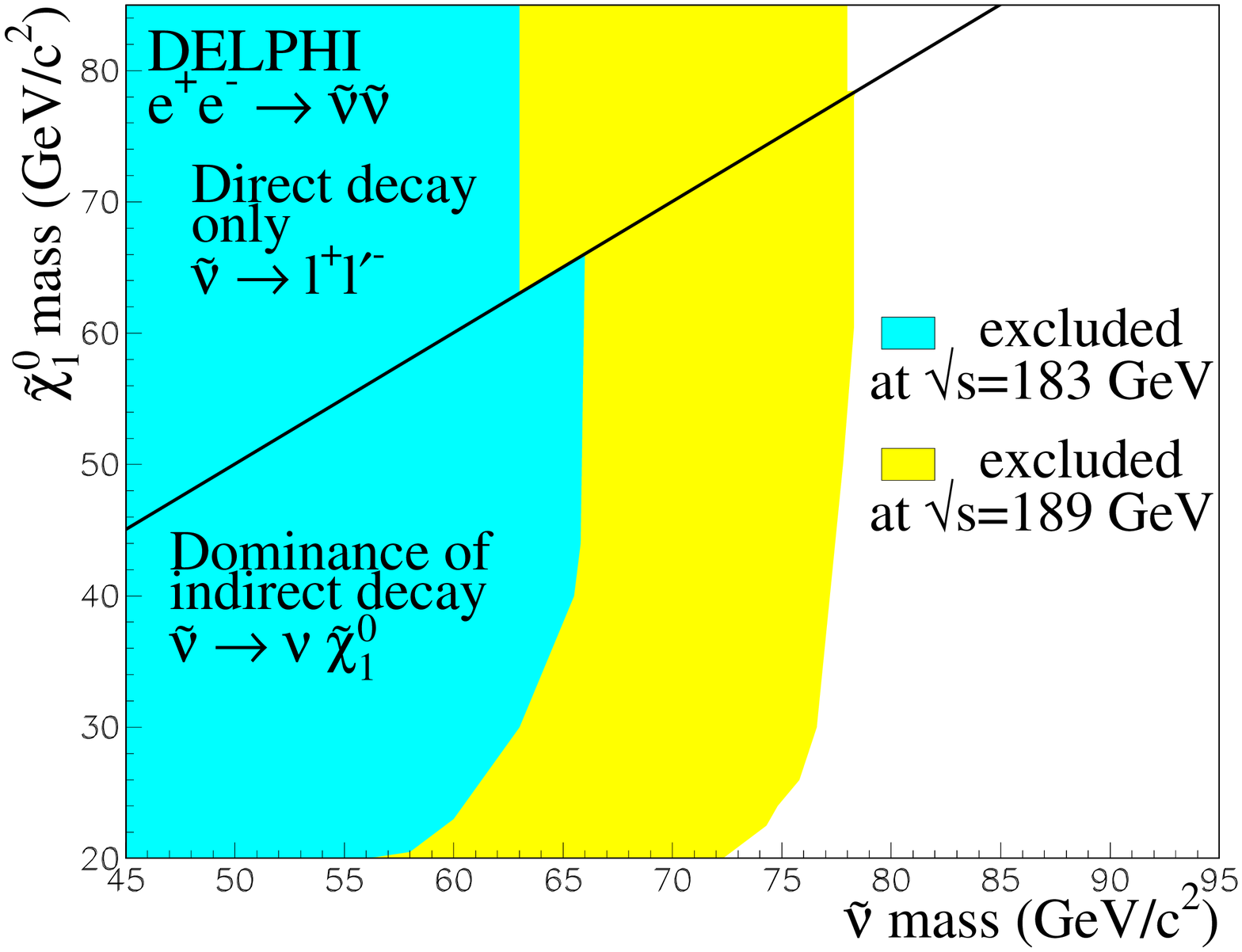,width=10cm}
\caption{
Excluded region  at 95\%~C.L. in m$_{\tilde{\chi}^0}$,
m$_{\tilde{\nu}}$ parameter space by \snu\ pair production
for direct and indirect decays.
The dark grey area shows the part excluded  by the
searches at 183~\GeV, the light grey area the one excluded by
the present analysis.}
\label{snuind}
\end{center}
\end{figure}

The results obtained from the sneutrino pair production studies
were used to derive limit on the sneutrino mass.
In the case of the sneutrino direct decay, the results improved
the upper limit on the sneutrino pair production cross-section.
Taking into account the results of the two analyses and
the efficiencies obtained when varying the sneutrino
mass, the cross-section limits for 2e2$\tau$ and  4$\tau$ channels were
derived and are reported in Fig.~\ref{snudir}.
The \snue\asnue\ cross-section
depends not only on the \snue\ mass but also on other MSSM
parameters (due to the possible $t-$channel \XPI\ exchange
contribution) and it is plotted for a specific
MSSM point: M$_2$~=~100~\GeVcc\
and $\mu$~=~--200~\GeVcc. The upper limit on the cross-section 
leads to a lower limit on the sneutrino mass of 78~\GeVcc.

\begin{figure}[h]
\begin{center}
\epsfig{file=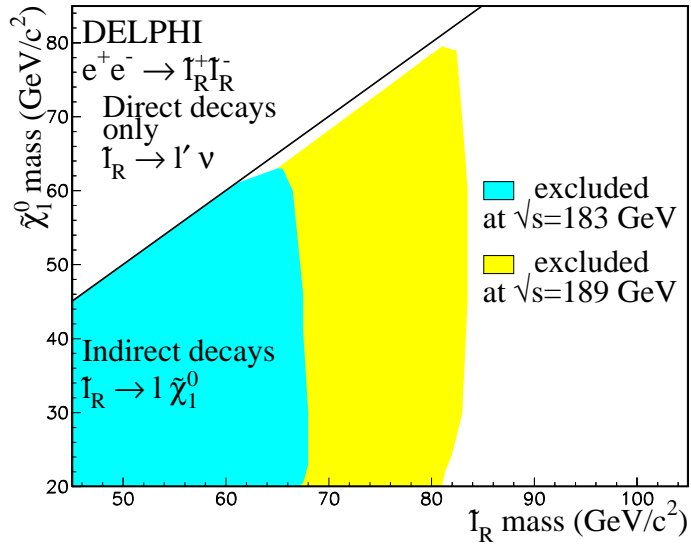,width=10cm}
\caption{
Charged slepton indirect decay:
excluded region  at 95\%~C.L. in
m$_{\tilde{\chi}^0}$, m$_{\tilde{\ell}_R}$  parameter space
by \slep$_R$ pair production.
The dark grey area shows the part excluded  by the
searches at 183~\GeV, the light grey area the one excluded by
the present searches at 189~\GeV.}
\label{slpind}
\end{center}
\end{figure}

In the case of the \snu\ indirect decay into $\nu$\XOI\ with
\mbox{the \rpv} decay of the neutralino via \Lacc, 
the efficiencies depend on 
the sneutrino and neutralino masses. The search results
allowed an area  
in the m$_{\tilde{\chi}^0}$ versus m$_{\tilde{\nu}}$ 
plane to be excluded, as shown on Fig.~\ref{snuind}. 
The same procedure has been followed for the charged slepton 
indirect decays.
The indirect decay of a \stau\ pair gives two taus and two
neutralinos,
and the final state selection was less efficient than 
for the \sel\ or \smu\ pair;
the results obtained for 
the \staur\ pair production gave the most
conservative limits on the slepton mass for any flavour, 
assuming that $\tilde{\ell}_R$ decays
exclusively to $\ell$\XOI. 
The area  excluded  in the
m$_{\tilde{\chi}^0}$ versus m$_{\tilde{\ell}_R}$ 
plane is plotted in Fig.~\ref{slpind}. The region where
m$_{\tilde{\ell}_R}$~-~m$_{\tilde{\chi}^0}$ is less than 2--3~\GeVcc\
was not covered by the present analysis, 
since then the direct decay
becomes the dominant mode, leading to two leptons and
missing energy.
Taking into account the limit on the neutralino mass at 30~\GeVcc, 
sneutrinos with mass lower than 76.5~\GeVcc\ and 
supersymmetric partners of the right-handed lepton,  
decaying indirectly, with mass lower than 83~\GeVcc\ were
excluded at 95\% C.L.

\section{Summary}

Searches for $R$--parity violating effects in e$^+$e$^-$ 
collisions at $\sqrt{s}=$~189~GeV
have been performed with the DELPHI detector.
The pair productions of neutralinos, charginos and 
sleptons have  been studied under the assumption 
that the $\rm{LL} \bar{\rm E}$ term is responsible for 
the supersymmetric particle decays into standard particles. 
It was assumed that one \Lijk\ coupling is dominant at a time
and that the
\Lijk\ coupling is strong enough for the LSP to decay within a few centimetres.
No evidence for $R$--parity violation has been observed, allowing
to update the limits previously obtained at $\sqrt{s} = $183~\GeV.
The present 95\%~C.L. limits on supersymmetric particle masses are:
\begin{malist}
\item $m_{\tilde{\chi}^0}>~$30~\GeVcc\ and $m_{\tilde{\chi}^\pm}>$~94~\GeVcc;
\item $m_{\tilde{\nu}}>$~76.5~\GeVcc\ (direct and indirect decays);
\item $m_{\tilde{\ell}_R}>$~83~\GeVcc\ (indirect decay only).
\end{malist}
These limits are valid for tan$\beta \geq $~1 and m$_0 <$~500~\GeVcc\ and for
all the generation indices $i$,$j$,$k$ of the $\lambda_{ijk}$
coupling, and for any coupling value 
from 10$^{-4}$ up to the existing limits.
%


\end{document}